\documentclass[twocolumn]{article}
\usepackage[a4paper, total={6in, 8in}]{geometry}
\usepackage[utf8]{inputenc}

\usepackage{todonotes}
\usepackage{graphicx}
\usepackage{float}
\usepackage{stfloats}
\usepackage{hyperref}
\usepackage{cite}

\title{Improving detection of protein-ligand binding sites with 3D segmentation.}
\author{%
  Marta M. Stepniewska-Dziubinska\textsuperscript{1}%
  \and Piotr Zielenkiewicz\textsuperscript{1,2}%
  \and Pawel Siedlecki\textsuperscript{1,2,*}%
  }
\date{%
\small
    \textsuperscript{1}Institute of Biochemistry and Biophysics, Polish Academy of Sciences, Pawinskiego 5a, 02-106 Warsaw, Poland\\%
    \textsuperscript{2}Department of Systems Biology, Institute of Experimental Plant Biology and Biotechnology, University of Warsaw, Miecznikowa 1, 02-096 Warsaw, Poland\\[2ex]%
    \textsuperscript{*}pawel@ibb.waw.pl
}

\begin{document}
\maketitle

\begin{abstract}
In recent years machine learning (ML) took bio- and cheminformatics fields by storm, providing new solutions for a vast repertoire of problems related to protein sequence, structure, and interactions analysis.
ML techniques, deep neural networks especially, were proven more effective than classical models for tasks like predicting binding affinity for molecular complex.

In this work we investigated the earlier stage of drug discovery process~--~finding druggable pockets on protein surface, that can be later used to design active molecules.
For this purpose we developed a 3D fully convolutional neural network capable of binding site segmentation.
Our solution has high prediction accuracy and provides intuitive representations of the results, which makes it easy to incorporate into drug discovery projects.
The model's source code, together with scripts for most common use-cases is freely available at \href{http://gitlab.com/cheminfIBB/kalasanty}{http://gitlab.com/cheminfIBB/kalasanty}.
\end{abstract}

\section{Introduction}
The aim of rational drug design is to discover new drugs faster and cheaper.
Much of the effort is put into improving docking and scoring methodologies.
However, most techniques assume that the exact location of binding sites~--~also referred to as \emph{pockets} or \emph{binding cavities}~--~is known.
Such pockets can be located both on a surface of a single protein (and be used to modulate its activity) or at protein-protein interaction (PPI) interfaces (and be used to disrupt the interaction).
This task is very challenging and we lack a method that would predict binding sites with high accuracy -- most methods are able to detect only 30\%-40\% of pockets~\cite{benchmark,deepsite}.

Traditional approaches for binding cavity detection are typically geometry-based~\cite{pass,pocketpicker,fpocket,depth}, but there are also  examples of tools using binding energy to different chemical probes~\cite{ftsite,sitehound}, sequence conservation (template or evolutionary methods)~\cite{concavity,glosa,probis}, or a combination of these~\cite{ligsitecsc,metapocket}.
For example, ProBiS~\cite{probis}~--~similarity-based tool, uses local surface alignment with sub-residue precision, allowing to find sites with similar physicochemical properties to the templates stored in the database.
Such methods simultaneously detect binding sites and provide some insight into their expected properties~--~they are most probably similar to the templates they were matched to.
Other approaches rely on a two-step algorithm, in which potential pockets are first identified and then scored to select the most probable binding sites.
For example, Fpocket~\cite{fpocket} is a geometry-based method, which first finds cavities in a protein's structure and then scores them.
The reverse approach is used in P2RANK~\cite{p2rank}, which uses a random forest (RF) model to predict ``ligandibility'' score for each point on a protein's surface, to then cluster points with high scores.

The latter tool is an example of applying machine learning (ML) to detect pockets~--~supervised ML to score surface points and unsupervised ML to post-process these predictions.
Unsupervised ML models are trained on unlabeled observations and aim to find patterns in the data in order to simplify their representation, remove the noise, and get better understanding of their nature.
Supervised ML models, on the other hand, require the observations to be paired with their corresponding labels (expected output, class, etc.).
The main purpose of this class of models is finding the relationship between the data and the labels that are actually desired.
The data is relatively readily available (in case of P2RANK~--~the structure of a protein) but the desired information is typically much harder to acquire (e.g. location of binding sites).

Another axis of classification of ML models is based on their complexity, or \emph{depth}.
Deep learning (DL) is a branch of ML grouping more complex models of different types, both supervised and unsupervised.
There is no clear border between ``classical'' ML and DL, but in general deep models are complex, multilayer neural networks capable of finding more sophisticated and convoluted relations between input and labels than their shallow counterparts.
DL models require less manual work and feature engineering, and use model's internal layers to extract features from the unprocessed data.

In the context of bio- and cheminformatics DL allows to predict \emph{in silico} properties that require much effort to establish experimentally, like detecting functional motives in sequences~\cite{dl_seq} or assessing binding affinity for protein-ligand complexes~\cite{kdeep,pafnucy}.

A recent example of a deep model used for binding site detection is DeepSite~\cite{deepsite}.
Similarly to P2RANK, DeepSite classifies each point in a 3D space based on its local environment as belonging (or not) to a binding pocket.
Probabilities for all points form a 3D density, that can then be post-processed to get the most probable locations and shapes of binding sites present in the structure.
Unlike P2RANK however, DeepSite uses a deep 3D convolutional neural network, with an architecture typical for image classification problems~--~set of convolutional layers paired with max pooling layers, followed by fully connected layers and a final neuron with the predicted class.
Such an architecture allows for the extraction of features~--~3D structural patterns~--~that are immediately used by the model to make predictions.

DeepSite was proven superior to two other state-of-the-art approaches at the time: Fpocket~\cite{fpocket} and Concavity~\cite{concavity}. The first one is a geometry-based, whereas the second~--~a sequence-conservation-based method. But although DeepSite achieved better results, only approximately 50\% of the predicted pockets is at most 4\r{A} from the actual position of the binding site. This is not an acceptable standard and calls for an improvement.

In this work we present a different DL-based approach for finding binding pockets, inspired by semantic image segmentation instead of classification.
Image segmentation aims at locating an object, or multiple objects, in an image.
Output of such a model is a set of scores assigned to each pixel, where the score denotes the probability that a given pixel belongs to the desired object.

In our case, the input is a 3D structure of a protein represented with a grid that can be analyzed in the same manner as 3D images, whereas the desired object is the binding pocket.
Our model called Kalasanty is based on U-Net~\cite{unet}~--~a state of the art model for image segmentation.
We adapted this model to the problem of binding cavity detection, and added functionalities that allow to easily generate predictions for protein structures.
The model takes protein structure as input, automatically converts it to a 3D grid with features, and outputs probability density~--~each point in the 3D space has assigned probability of being a part of a pocket.
Predictions can then be saved as \texttt{.cmap} or \texttt{.cube} files, that can be later analyzed in molecular modeling software.
Kalasanty can also output parts of the protein that form pockets and save them as \texttt{.mol2} files.

\section{Methods}

\subsection{Our approach}

In order to solve any problem with DL, it first needs to be specified in terms of how input and output are represented.
In this work we formulate the pocket detection task as a 3D image segmentation problem.
This allowed us to use well established DL methods, originally developed for 2D images.

With this approach both input and output are represented as 3D grids with the exact same dimensions~--~each grid point in the input has a corresponding point in the output.
The input is a discretized protein structure with multiple feature channels describing atomic properties.
The returned output has a single channel with probability of belonging to the pocket
(details in section ``Data'').
Known pockets, which are used for training and evaluation, are represented with binary grids, where 1s denote grid points that belong to the pocket and 0s otherwise.

\begin{figure*}[hb!]
	\includegraphics[width=\textwidth]{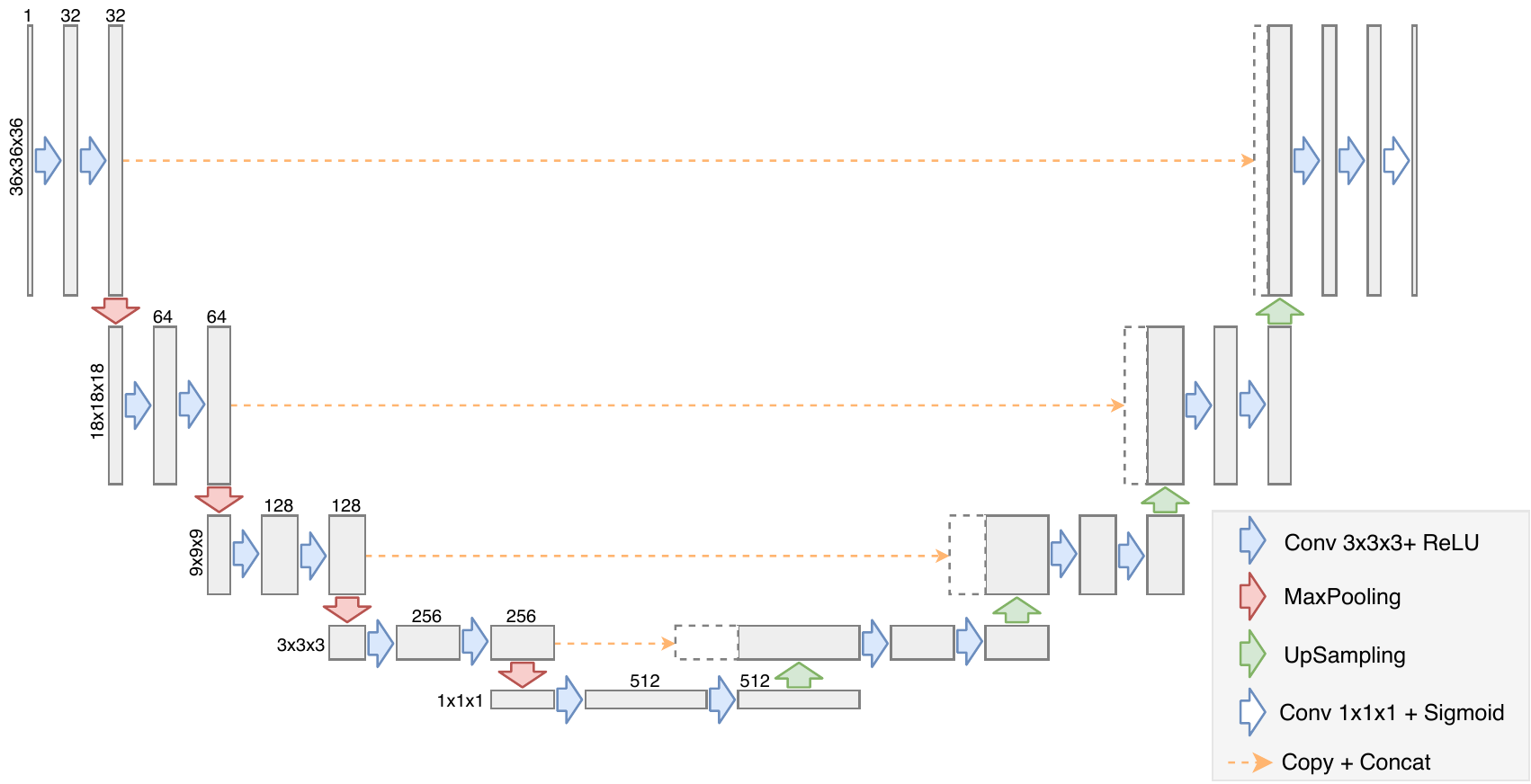}
    \caption{Kalasanty's architecture.}
    \label{fig:model}
\end{figure*}

The model is based on an architecture called U-Net (see Figure~\ref{fig:model}), which is a fully-convolutional, encoder-decoder model that pioneered skip connections (see text below).
This kind of architecture prevents from loosing fine-grained information about the input which greatly increases the precision of the resulting segmentations.

Kalasanty was built using the same ideas and architecture design as the original U-Net.
Similarly to the original network, it is composed of blocks consisting of two convolutional layers and a single max-pooling or up-sampling layer (depending on the side of the ``U'').
However, Kalasanty works on 3D data instead of 2D images, therefore 3D versions of convolutional, pooling and up-sampling layers were used.
Also, the number of layers, number of convolutional filters, and patch sizes were adjusted to match the size of the input and difficulty of the task.

Kalasanty has 9 convolutional blocks~--~4 in the encoder, one in the bottleneck, and 4 in the decoder part of the network.
Each block consists of two convolutional layers with the same number of filters (32, 64, 128, 256, or 512), kernel size of 3x3x3 pixels and ReLU activation function, combined either with a max-pooling layer (encoder path, left side of the Figure~\ref{fig:model}) or with an up-sampling layer (decoder path, right side of the Figure~\ref{fig:model}).
The two first max-pooling layers and the two last up-sampling layers have 2x2x2 patch sizes, while layers in the middle have 3x3x3 patch sizes.
This way, for input of 36x36x36 pixels that was used in this work, feature maps in the middle of the network (bottleneck, bottom of the Figure~\ref{fig:model}) have spacial sizes of 1x1x1 and can be used as feature vectors.

What is unique about U-Net-like models when compared with other encoder-decoder networks it that the information between the two paths is not only passed through a bottleneck, but also after each block using so-called skip connections.
The final feature map from each block in the encoder is copied and concatenated with the first feature map in the corresponding decoder block (orange dashed lines).
This allows to better localize features and therefore return more accurate segmentations.

The model was defined with the Keras library~\cite{keras}.
Apart from methods needed for ML-related tasks, we implemented custom methods for working with the molecular data: making predictions for molecules; locating amino acids forming the pockets; saving the predicted probabilities as \texttt{.cmap} or \texttt{.cube} files; and saving parts of proteins forming pockets as \texttt{.mol2} files.
Source code and network's parameters are freely available at \href{http://gitlab.com/cheminfIBB/kalasanty}{http://gitlab.com/cheminfIBB/kalasanty}.

Models were trained with L2 regularization on each layer's weights ($\lambda=10^{-5}$).
Also, random translations and rotations were used during training to augment the dataset.
As the objective function we used the negative Dice coefficient for continuous variables:
\[
C(y, t) = \frac{ -2 \sum_{i,j,k} (y_{i,j,k} \cdot t_{i,j,k}) + \epsilon}{\sum_{i,j,k} (y_{i,j,k} + t_{i,j,k}) + \epsilon}
\]
where $y$ and $t$ are the predicted and target segmentations, $(i,j,k)$ are indices of a grid cell, and $\epsilon$ is a smoothing factor ($\epsilon=0.01$ was used).
Minimization was performed for $1.5\cdot10^{6}$ steps with 10 samples in each batch, and the Adam optimizer with a learning rate of $10^{-6}$ and with default values for the remaining parameters~\cite{adam}.

\subsection{Data}\label{sec:met_data}
For training and validation of the model the sc-PDB~\cite{scpdb} dataset was used.
The database consists of known binding sites, accompanied with prepared protein structures.
Binding sites are represented both with 3D shapes of cavities generated with VolSite~\cite{volsite} (which were used in this study) and amino-acids that form them.
VolSite describes a binding cavity with a set of pharmacophoric properties arranged on a 3D grid, based on the properties of the neighboring protein atoms.
Data is stored as mol2 file, with atoms encoding each property.
In this project, we converted data in the grid to a binary information -- whether a point in space is part of a pocket or not.
Grid cell (see section ``Our approach'') with such point inside was considered a part of the pocket.

The database (v 2017) consists of 17594 binding sites, corresponding to 16612 PDB structures and 5540 UniProt IDs.
The test set was constructed from a different dataset, one used for benchmarking by Chen \emph{et~al.}~\cite{benchmark} (see below).
When assessing the quality of a ML model it is important to evaluate it using a separate dataset.
Using a different dataset minimizes the risk that there were some database-related artifacts, that might have been exploited by the model.
However, it is easy to make a mistake and have the same protein (with slightly different structures) in the training and the test set as well.
This is a common error called \emph{data leakage} which leads to overly optimistic assessment of a model.
In order to avoid data leakage, all structures of proteins from the benchmark were removed from the sc-PDB database (481 structures).
Also, 304 binding sites were discarded because of errors when loading their corresponding protein structures with Open Babel.
Finally, 15860 structures, corresponding to 5473 UniProt entries, were used for training.

Number of structures per protein varied from 1 to 280, with median equal to 1 and mean equal to 3.26.
The dataset contained protein structures originating from 952 different organisms, from which the most abundant were human (34.4\%), \emph{E. coli} (5.6\%), Human immunodeficiency virus (4.2\%), rat (2.9\%), and mouse (2.4\%).
Also, diverse protein architectures were well represented, with 5171 structures of mainly alpha proteins, 2500 of mainly beta proteins, 11758 of alpha-beta proteins and 53 structures of proteins with few secondary structures~\cite{cath}.
The dataset was also diverse from a sequence perspective and contained proteins from 1983 different Pfam families and 982 superfamilies, protein kinases being the most frequent.

The data from sc-PDB were split into 10 folds (subgroups containing 1586 structures each) based on UniProt ID, i.e. all structures of a single protein must be in the same fold.
This setup was necessary to avoid data leakage during validation.
Putting different structures of the same protein in different folds would result in having almost identical training and validation examples, which might result in unnoticed overfitting, and as a consequence overoptimistic evaluation and possibly selection of incorrect hyperparameters.

We also analysed binding site similarity across obtained folds (using Shaper~\cite{volsite} results provided by sc-PDB) to assure that there is no data leakage during validation.
This setup was used to train the models with 10-fold cross-validation (CV).
CV results were used to select model and optimization parameters and assess models' stability.
The final model was trained on all 10 folds combined to achieve the best possible performance.

As mentioned, for the test set we used structures from the Chen benchmark~\cite{benchmark}.
This benchmark set contains apo and holo structures for 104 proteins (208 in total).
Structures were converted to the format used in sc-PDB to evaluate the models using the following steps.
First, for each structure ligand(s) and protein were split into separate files using UCSF Chimera~\cite{chimera}.
Solvent and ions were assigned to the protein.

Then, we used VolSite~\cite{volsite} (available in IChem toolkit) to describe a cavity for each ligand.
In case of apo structures, they were aligned to their holo counterparts, and then the ligands were used to select pockets.

For 59 structures (31 apo and 28 holo) out of the 208 present in the benchmark, VolSite failed to find a cavity because of its insufficient buriedness.
Still, the remaining part of the dataset (149 structures with 269 binding sites) offers a valuable test set because it contains diverse proteins not used for training.

We used the final 149 structures with 269 binding sites to assess the performance of our model and the performance of the DeepSite model.
Note, that we have no control over the dataset that was used by \cite{deepsite} and 12 proteins from the test set have been used to train DeepSite.
This might result in a slightly more optimistic evaluation for DeepSite, but we have decided to keep those structures so that the test set was larger.

It is important to note, that the results presented in this work cannot be directly compared to the ones presented by Chen \emph{et~al.}~\cite{benchmark} because only a subset of the original dataset was used.

Finally, all the resulting protein structures and segmentations were represented with 3D grids with 2\r{A} resolution.
The grids were centered on a protein center and had 70\r{A} in each direction.
Proteins were described with 18 atomic features used in our previous project~\cite{pafnucy}.
Pockets were transformed to 3D binary masks with same size, center, and resolutions as grids representing their corresponding protein structures.

For 79 (0.5\%) entries in the sc-PDB database this procedure lead to empty pocket grids.
After manually inspecting several of such cases it turned out that it affects large protein complexes and less carefully prepared protein structures (e.g. 1zis, which contains two unbound protein chains, located far away from each other).
Although such structures may arise in high-throughput experiments and analyses (like the one presented in this study), they are highly unlikely to occur in real-life studies aiming to discover binding cavities in a protein of interest.
Also, this issue affects a neglectable fraction of the training data and none of the test examples.
We therefore decided not to modify the structures, nor the procedure for data preparation.
The data with empty pocket grids were used for training as negative examples, and skipped in the validation.

\subsection{Results evaluation}
Results obtained with Kalasanty and DeepSite were evaluated using two popular metrics: $DCC$ and $DVO$ (discretized volume overlap).
$DCC$ is the distance between the predicted and the actual center of the pocket.
It is typically used to describe a success rate for the method, i.e. the fraction of sites below the given $DCC$ threshold.
Similarly to other authors~\cite{benchmark,deepsite,pocketpicker}, we analyzed success rates for thresholds up to 20\r{A} and considered pockets with $DCC$ below 4\r{A} as correctly located.
$DVO$, on the other hand, is a more strict metric comparing shapes of the predicted and actual pockets.
It is the volume of the intersection of the predicted and the actual segmentations, divided by volume of their union.
The two metrics complement one another, highlighting different aspects of prediction quality~--~correct location ($DCC$) and shape ($DVO$) of predicted pockets.
When used together, they provide concise yet rich description of the results, and allow to analyze them faster and more objectively than with visual investigation.

In order to calculate both metrics, predicted densities were converted to binary segmentations.
For Kalasanty, probability threshold of 0.5 was used, which was selected based on models' performance on the validation set.
However, we note that using thresholds between 0.3 and 0.8 leads to very similar results.

DeepSite predictions were obtained using the playmolecule.org web-service.
Predicted binding cavities were download as \texttt{.cube} files and converted to binary masks using the probability threshold of 0.4, which was recommended by DeepSite's authors.

Following the work of Chen \emph{et~al.}~\cite{benchmark}, for each structure only $n$ predicted pockets with the highest scores were considered, where $n$ is the number of pockets present in the structure.
Then, each predicted pocket was matched with the closest real pocket and $DCC$ and $DVO$ values were calculated.
If no pocket was predicted, we used the worst possible values for the metrics, which were $DVO=0$ (no overlap) and $DCC=70\sqrt{3}$\r{A}$=121.24$\r{A} (the biggest possible distance for a 70\r{A} cube).

Additionally we used F1 score~--~metric typically used to evaluate ML models for detection tasks.
F1 score combines precision (positive predictive value) and recall (also called sensitivity)~--~its their harmonic mean.
Conversely to other popular metrics used in ML, like accuracy or ROC AUC, F1 do not require notion of true negative, which is undefined for detection problems.
We used $DCC$ of 4\r{A} as a threshold for true positives when calculating these metrics.

\section{Results}

In this study we present two sets of results.
In the first part, we describe cross-validation (CV) experiments, in which we tested Kalasanty's stability and its general properties, using nearly 16k structures from the sc-PDB database.
In the second part, we show results for the model trained on the whole training set and evaluated on the test set.
We also compare it to another DL-based approach~--~DeepSite.

For CV experiments, the training set was divided into 10 parts.
Then, 10 models were trained with one fold left out for validation.
This way we were able to evaluate our approach on the whole dataset, without making predictions for structures that were used to train a particular model.

\begin{figure}[hbt!]
	\centering
	\includegraphics[width=0.5\textwidth]{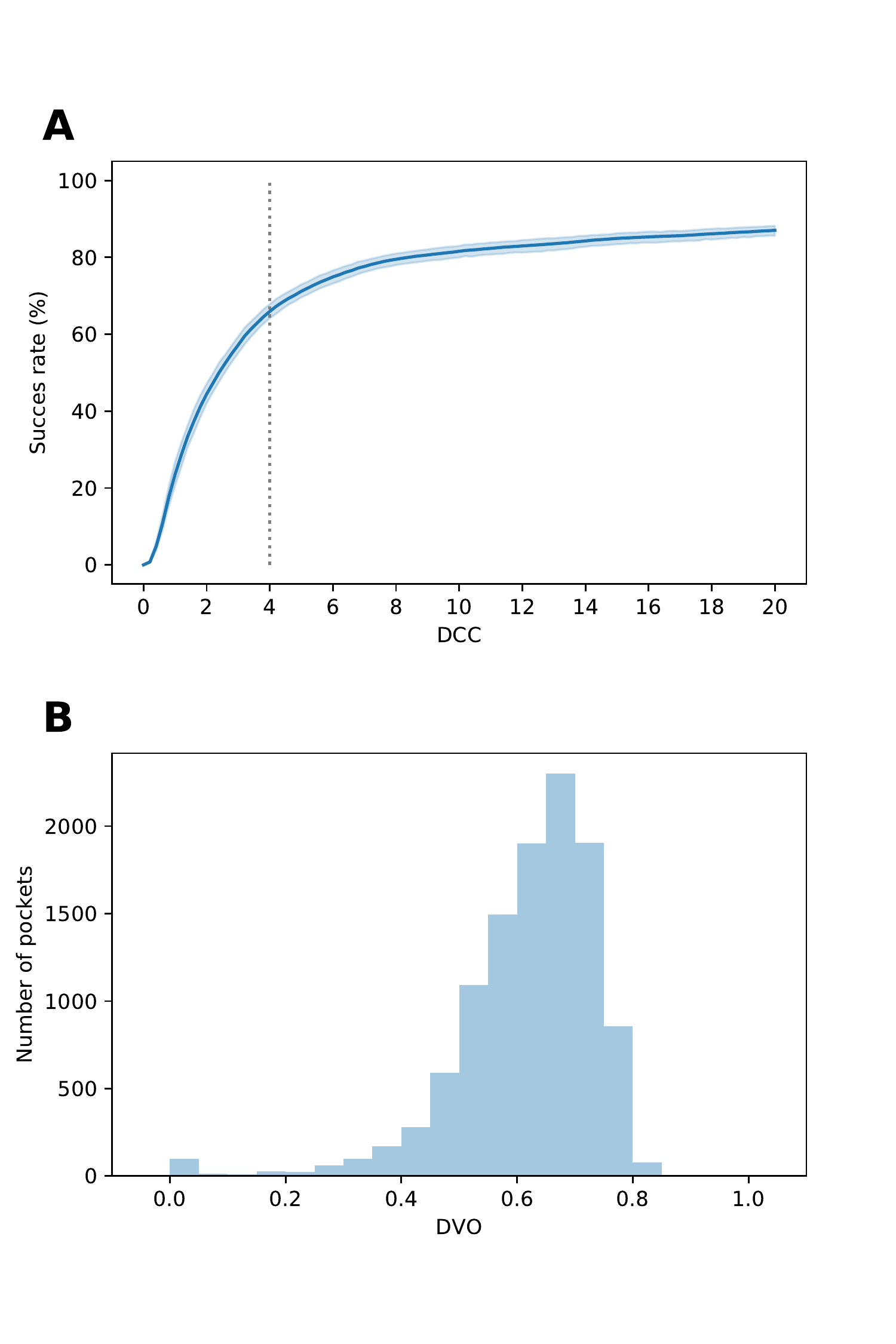}
    \caption{
    Models' performances on the validation sets. A) Success rate plot for different $DCC$ thresholds, averaged over 10 folds. Blue area around the curve depicts 95\% confidence interval. 4\r{A} threshold is marked with gray dotted line. B) $DVO$ distribution for correctly located pockets ($DCC < 4$\r{A}) in all folds combined.
    }
    \label{fig:cv_scores}
\end{figure}

Also, the validation sets were used to monitor the training and to select optimal parameters defining the network architecture and optimization procedure.
For each model we observed a plateau on the learning curve for the validation set.
Results were stable across folds and we observed similar distributions of $DCC$ values for each fold (see Figure~\ref{fig:cv_scores}).
The variation for $DCC$ is so small, that 95\% confidence interval for mean estimation almost melts into the mean curve (Figure~\ref{fig:cv_scores}A).
The model did not return any pockets for 4.7\% of the structures (737 pockets, ranging from 31 in fold 6 to 113 in fold 2).
F1 score was equal to 0.69, with recall of 0.66 and precision of 0.73.

We have also looked for global trends in prediction accuracy and factors, that should be irrelevant for the model in order for it to generalize well.
We had an unbalanced dataset and the model might have been biased in favor of the most prevalent types of proteins.
Although the performance differs between different groups of proteins (see the ``Discussion''), we did not observe any systemic differences between results for the most frequent types of proteins and average results.
This also suggests that the model did not overfit to the training set, and that it generalizes well to new, unseen structures.

Next, we used the best set of parameters and trained the final model on the whole sc-PDB dataset (15860 structures).
Finally, we evaluated the performance of this model and DeepSite's on the test set.
This was a more challenging dataset, containing proteins from a different source.

\begin{figure*}[ht!]
    \centering
	\includegraphics[width=\textwidth]{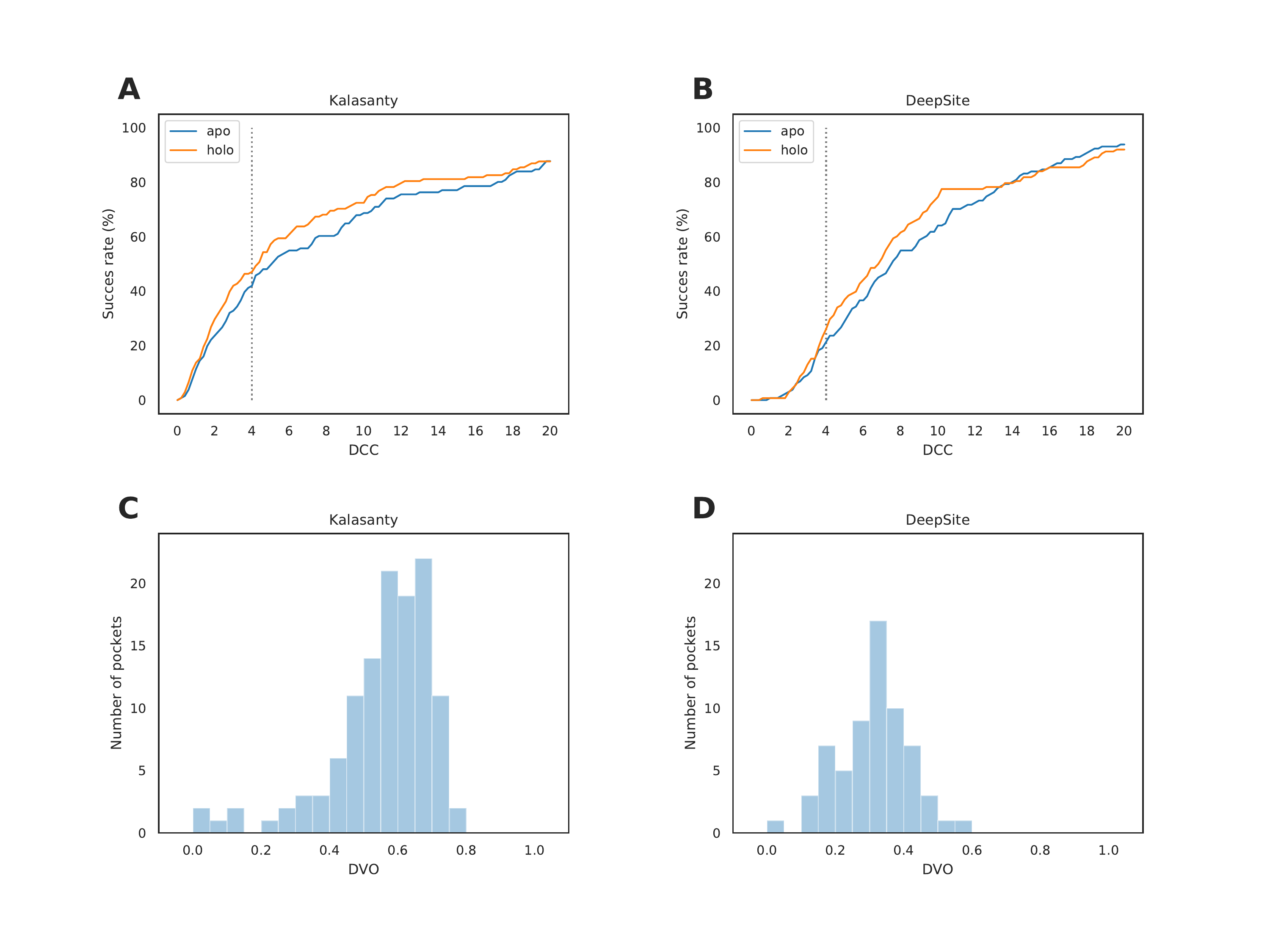}
    \caption{Models' performance on the test set. A) and B): Success rate plot for different $DCC$ thresholds for Kalasanty and DeepSite, respectively. 4\r{A} threshold is marked with gray dotted line. C) and D): $DVO$ distribution for correctly located pockets ($DCC < 4$\r{A}) for Kalasanty and DeepSite, respectively.}
    \label{fig:test_scores}
\end{figure*}

As expected, the performance was slightly worse than the one observed in CV experiments yet still promising (see Figure~\ref{fig:test_scores}, panels A and C).
For only 3 structures (1.3\%) no pockets were detected and for 120 of the binding sites (44.6\%) center of the predicted pocket was at most 4\r{A} from the center of the real binding site.
However, 5\% of correctly located pockets (compared to 1.5\% in the validation set) had incorrectly predicted shape, resulting in $DVO$ below 0.25.
The F1 score was equal to 0.45 (precision=0.64, recall=0.35).

DeepSite performance on the test set was also worse than CV results reported in \cite{deepsite} (see Figure~\ref{fig:test_scores}, panels B and D).
When threshold of 0.4 (recommended by the authors) was used, pockets were detected in all structures, but only 64 (23.8\%) of them had $DCC$ below 4\r{A}.
What is more, only 2 pockets in the entire test set had $DVO$ above 0.5.
It is worth noting that similarly to DeepSite's authors we did not observe significant changes in the results when different thresholds between 0.3 and 0.6 were used.
The overall F1 score was equal to 0.26 with precision of 0.36 and recall of 0.20.

\section{Discussion}
\begin{figure*}[ht!]
	\centering
	\includegraphics[width=\textwidth]{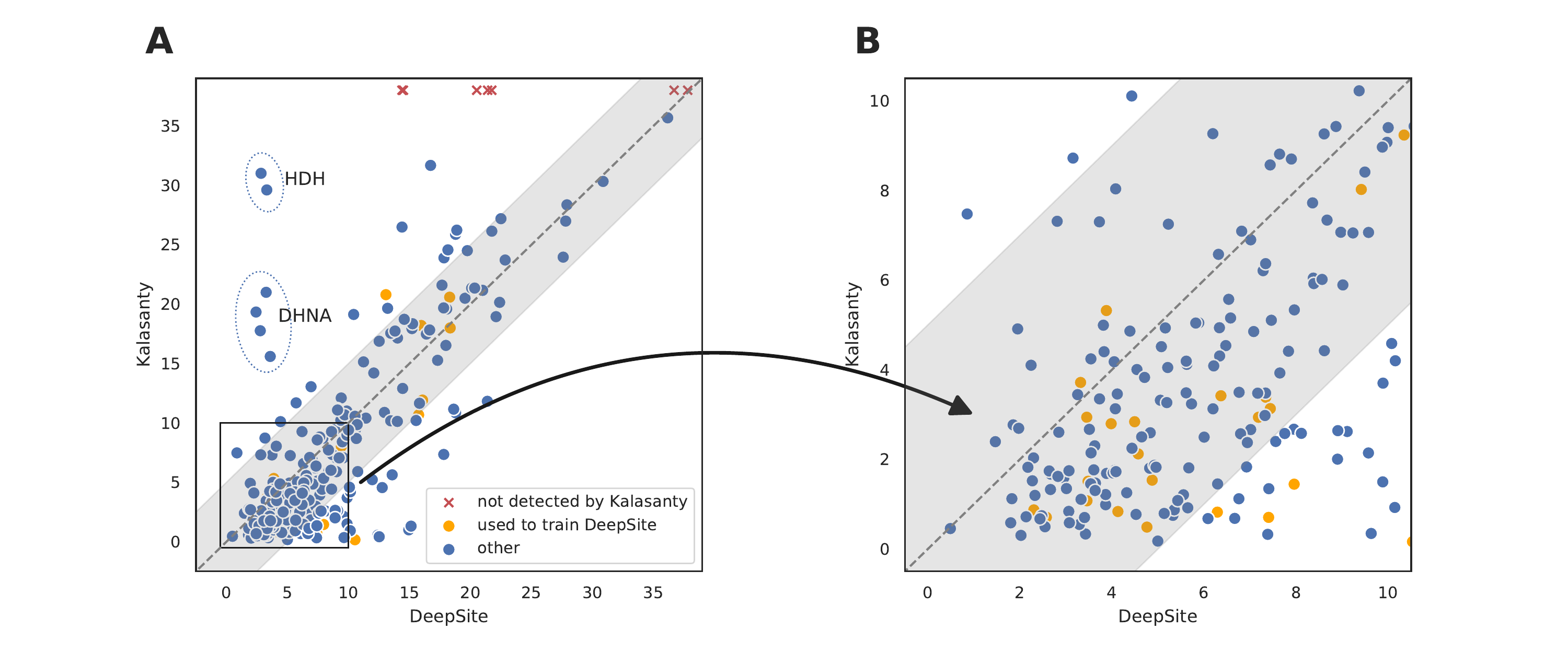}
    \caption{Comparison between $DCC$ values achieved by DeepSite and our model: A) on the whole test set and B) zoomed-in on high-quality predictions. Each point corresponds to $DCC$ values obtained with the two models. Binding sites for which U-Net did not return predictions are depicted with red crosses at the top of the image. Structures that were used to train DeepSite are colored with orange. Gray area marks 5\r{A} difference in both directions from the diagonal (same predictions for the two models). Predictions for L-histidinol dehydrogenase and 7,8-dihydroneopterin aldolase are marked with ellipses and annotated with proteins' short names (HDH and DHNA, respectively).}
    \label{fig:scores_comp}
\end{figure*}

In order to better understand the difference between Kalasanty and DeepSite, we analyzed if the two models make similar mistakes (see Figure~\ref{fig:scores_comp}).
Although the general trends are similar and same proteins were problematic for the two models, Kalasanty correctly detected almost twice as many pockets as DeepSite (44.6\% vs 23.8\%, respectively).
Also, for 84.6\% (115 out of 136) of pockets detected by at least one of the models, Kalasanty had lower $DCC$ than DeepSite.

\begin{figure*}[ht!]
	\centering
	\includegraphics[width=\textwidth]{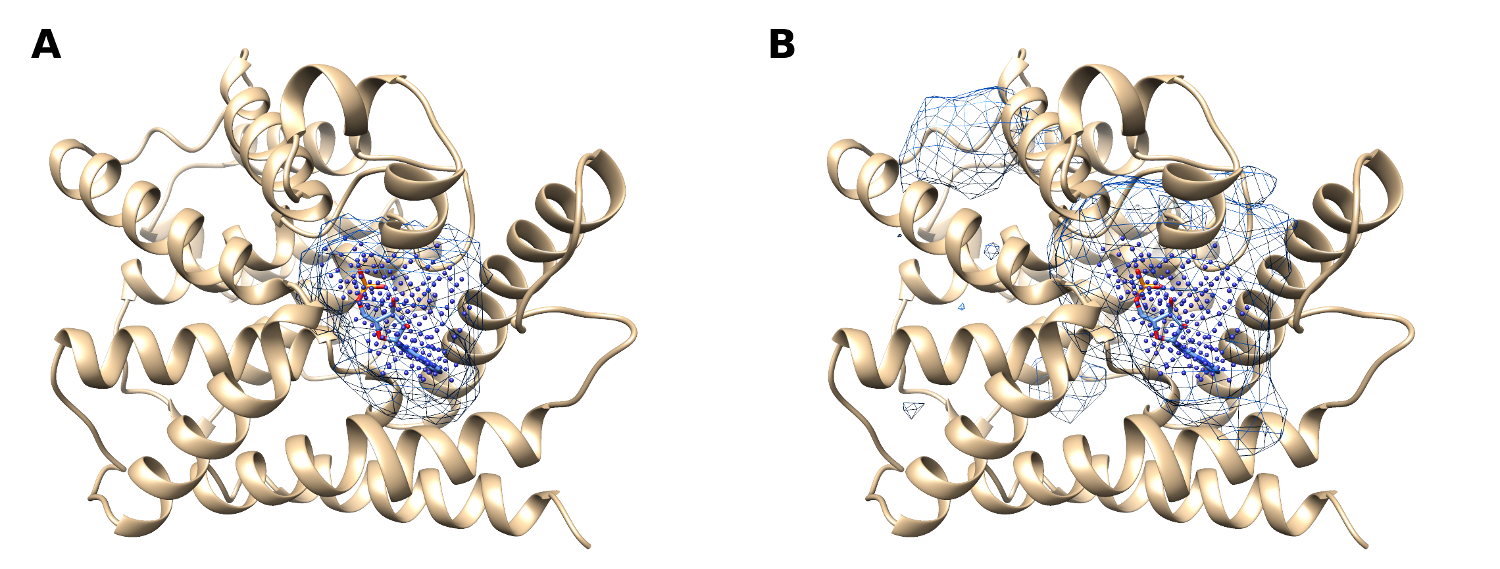}
    \caption{Binding site probability density for cAMP-specific 3',5'-cyclic phosphodiesterase 4D (UniProt ID: Q08499, PDB ID: 1tb7) predicted with A) Kalasanty and B) DeepSite. VolSite representation of the binding site (ground truth) is depicted with points, while prediction with a mesh.}
    \label{fig:viz}
\end{figure*}

This difference in performance is probably caused by the fact, that DeepSite tends to return more voluminous predictions (Figure~\ref{fig:viz}).
Although there is no single definition of a binding site (should it be a group of amino-acids, or a void between them?), the two models were trained on the same dataset so this comparison is justified.
After converting densities into binary predictions, pockets returned by DeepSite are on average twice as big as those predicted by our model.
It also explains the discrepancy between $DCC$ and $DVO$ results for DeepSite~--~even correctly located pockets are usually too big and their shape is not modeled accurately.

Interestingly, for two proteins~--~L-histidinol dehydrogenase (UniProt ID: P06988, PDB IDs: 1k75 and 1kae) and 7,8-dihydroneopterin aldolase (UniProt ID: P56740, PDB IDs: 1dhn and 2nm2)~--~DeepSite correctly located all pockets in both holo and apo structures, while Kalasanty completely missed them (see Figure~\ref{fig:scores_comp}A).

We investigated what is the root of the observed differences in those two specific cases. 
We were interested whether poor results for the two proteins are a part of a bigger trend and can be explained by our model's inability to predict pockets for a particular group of proteins. 
Unfortunately such analysis is hampered by the test set size which is too small to perform statistical analysis.
We therefore analyzed relationships between protein properties (superfamily, fold, source organism, length and size of the binding site) and prediction quality using CV results.
From these experiments we indeed can observe higher $DCC$ values for both superfamilies when compared to the rest of the dataset (Supplementary figure S1, panels A and B).
We also observed significant differences for both source organisms (Supplementary figure S1, panels C and D).
However, we cannot determine if those are causal relationships, or is there some other underlying factor, correlated with these two properties.

To summarize, we hypothesize that poor results for the two proteins might be related to some systematic errors in predictions made for such types of proteins.

\section{Conclusions}
In this work we presented Kalasanty~--~a neural network model for detecting binding cavities on protein surfaces.
Kalasanty was trained and validated with the sc-PDB database and additionally evaluated on an independent test set.
We compared Kalasanty with DeepSite which was proven better than Fpocket and Concavity~--~one of the best conventional methods for binding site prediction.
Results show that our model achieves high accuracy and is able to locate pockets more precisely than DeepSite (44.6\% and 23.8\% correctly located pockets from the test set, respectively).
What is also important, Kalasanty is stable and cross-validation results are comparable to those obtained for new data (not used for training nor validation).

However, it should be noted that the sc-PDB dataset contains only deep cavities, which tend to have better properties (druggability).
Model trained on such a dataset is not able to detect binding sites that are located on flat surfaces.
To obtain such a model, different datasets should be acquired which is out of the scope of this study.
]
Kalasanty is able to find multiple binding sites for a single protein.
However, if they are closely located the current post-processing procedure might merge them into one pocket.
A possible extension of this approach would be to look for local probability maxima around which candidate binding sites would be constructed.

Kalasanty was implemented in Python and the architecture was defined using the Keras library.
Source code, together with trained model and helper scripts are freely available at \href{http://gitlab.com/cheminfIBB/kalasanty}{http://gitlab.com/cheminfIBB/kalasanty}.
The repository can be also used to launch online demo, allowing to test Kalasanty through the web browser on a molecule of interest, without the necessity of installation nor registration.
The network was supplemented with additional methods that allow for making predictions directly for molecules, and handle all necessary preprocessing under the hood.
Predictions can then be saved as \texttt{.cmap} or \texttt{.cube} file and visualized in molecular modelling software.

Although Kalasanty is a deep neural network, using it does not require GPU.
GPU is crucial for training, but not for inference.
It takes 5 seconds to load the model and a second to make a prediction on a Intel Core i7 CPU.
This makes Kalasanty accessible for all researchers.

Kalasanty is based on U-Net~--~ a state-of-the-art neural network architecture for semantic segmentation, originally developed for 2D medical images.
We adapted the U-Net to process 3D protein structures and provided the model with input relevant for the task of identifying binding cavities.

Deep learning methods gained popularity in the recent years because of their flexibility and potential for capturing complex relationships hidden in the data.
The field of deep learning is ripe with noteworthy ideas that have already been tested in disciplines such as computer vision and sequence modelling.
Therefore, this work can also be seen as an example of adapting deep learning methods developed in other fields to structural bioinformatics.

\bibliographystyle{unsrt}
\bibliography{references}

\appendix
\renewcommand\thefigure{S\arabic{figure}}
\setcounter{figure}{0}

\onecolumn
\cleardoublepage
\section*{Supplementary Figures}

\begin{figure}[H]
	\centering
	\includegraphics[width=0.5\textwidth]{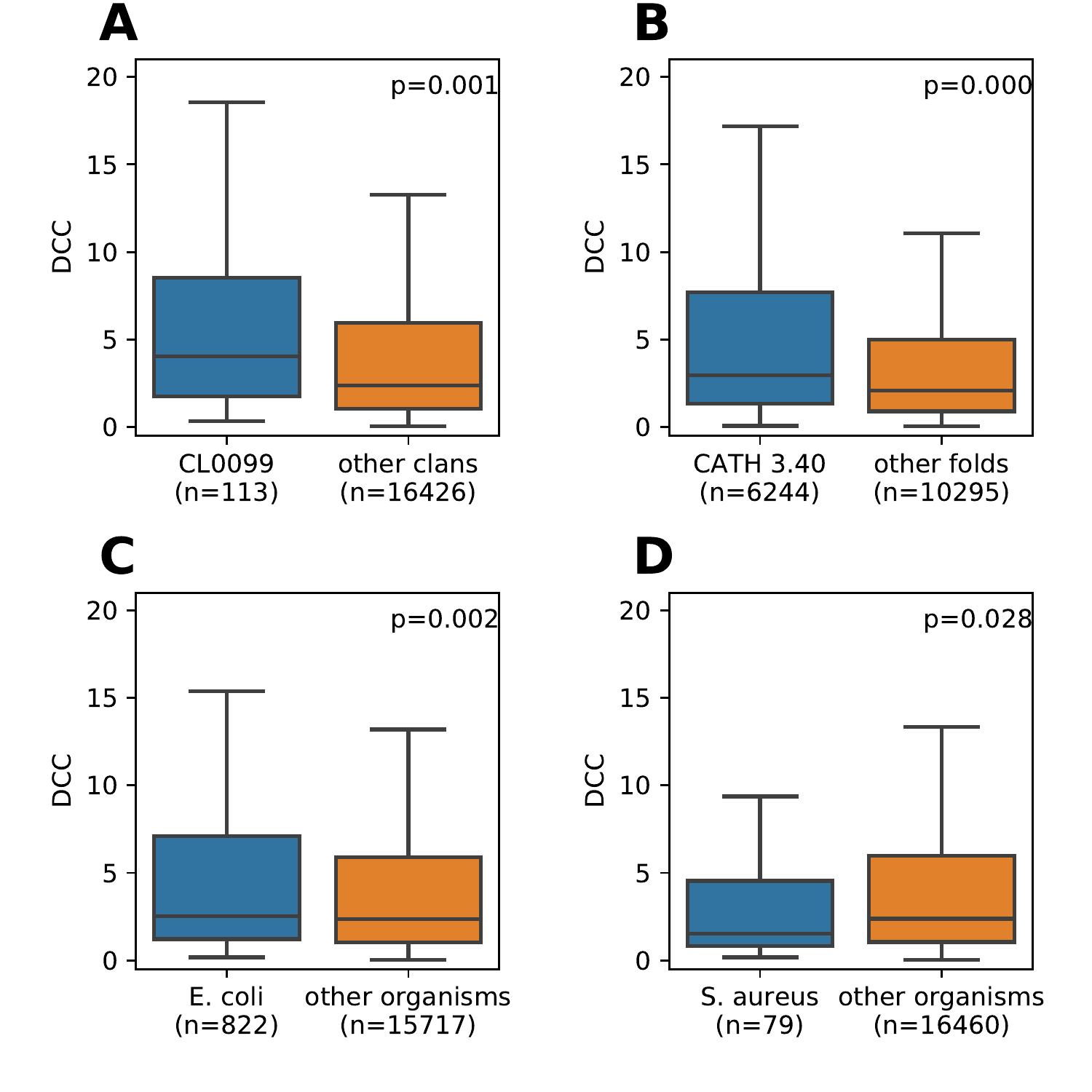}
    \caption{Relationship between prediction accuracy and protein properties: superfamily (A), fold (B), and source organism (C and D). ``n'' denotes number of binding sites in each group.
p-values were calculated with Mann-Whitney U-test.}
    \label{fig:metadata}
\end{figure}

\end{document}